# SIZE EFFECT ACCORDING TO THE CRITICAL SHEAR CRACK THEORY (CSCT) FOR REINFORCED CONCRETE BEAMS AND SLABS

REPLY TO THE SEGIM REPORT NO. 18-107788C BY A. A. DÖNMEZ AND Z. P. BAŽANT

**Aurelio Muttoni and Miguel Fernández Ruiz**

Ecole Polytechnique Fédérale de Lausanne (Switzerland)

## Introduction

This document is a reply to the SEGIM report No. 18-10/788c "Critique of Critical Shear Crack theory (CSCT) for *fib* Model Code Articles on Shear Strength and Size Effect of RC Beams" [1] by A. A. Dönmez and Z. P. Bažant dated 31.10.2018.

In that document, Dönmez and Bažant raise a number of criticism on the hypotheses Critical Shear Crack Theory (CSCT). The aspects criticized have however been largely discussed in previous works of the CSCT, proving the validity of the theory. These works do not seem to have been read or understood by the authors of [1] and this document compiles this previous knowledge as an explanation to the authors of the report.

In addition, some criticism is raised in [1] on the applicability of the CSCT to shear and punching shear design in codes of practice. The CSCT has shown to be general, reliable and simple enough to be introduced in codes of practice. This theory has been adopted by different codes (fib MC2010, Swiss Code, current draft of Eurocode 2) after in-depth analysis of different scientific committees and extensive verification. Nevertheless, the raised criticism is also addressed, highlighting the benefits of the implementation of the CSCT in codes of practice, as well as the deficiencies of the approach suggested in [1].

## The approach of the CSCT

The Critical Shear Crack theory (CSCT) was developed following a scientifically-consistent approach allowing for understanding and reproducibility of results. The theory is developed with a clear link to experimental observations, explaining the observed phenomena and calculating their response (extensive details for shear design can for instance be found in [2-3]):

1. The CSCT hypotheses are based and justified on experimental findings extensively published and subjected to the peer-review of the scientific community. Realistic tests performed with refined measurements (see for instance [3-4], an instance is presented in Fig. 1a for a DIC shear test on continuous members and Fig. 1b for the measured kinematics at the critical shear crack).
2. The experimental measurements are used in conjunction with well-established constitutive material laws (Hordijk, Walraven…) so to determine in an objective manner the stresses at the critical shear crack (see for instance [3-4] and Figs. 1c-d) and the amount of shear carried by all the potential shear-transfer actions. Other independent teams of researchers (see for instance [5]) have performed similar procedures confirming the consistency of such measurements and interpretation.

3. The amount of shear carried by the potential shear-resisting actions is tracked through the whole loading process (see Fig. 1e,f accounting for aggregate interlock, residual tensile strength, doweling action and inclination of compression chord).
4. Understanding and mechanical modelling is proposed on this basis [2-4]. Simplifications are introduced in the expressions to allow to expressions usable for design [6] but showing the physics of the problem.
5. The results are extensively validated with test results from own experimental programmes and from the literature (validations performed by the authors of the CSCT and independent teams [6]). The results show that not only the failure load is suitably reproduced, but also the amount of shear carried by each shear-transfer action (see Fig. 2,3). Also, the consistency of the expressions is systematically checked with known physical boundaries, as for the case of the size effect [7].

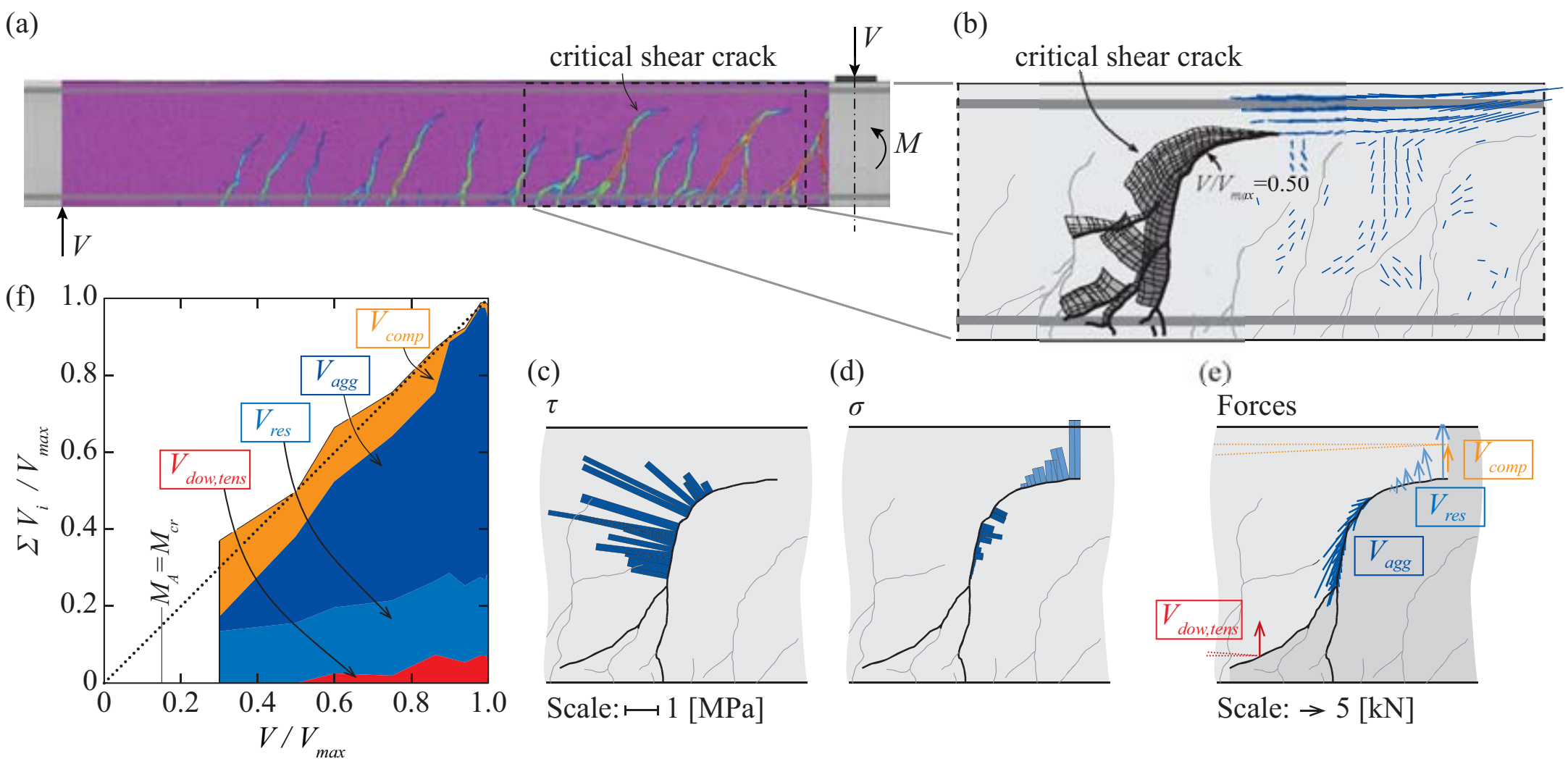

*Figure 1:* *The CSCT approach for shear (figure adapted from [3]): (a) testing and DIC measurements; (b) analysis of critical shear crack kinematics and principal strains in the compressive zone; (c-e) calculation of the crack stresses and interface forces on the basis of the measured kinematics and constitutive laws; and (f) tracking of shear-transfer actions contributions during loading process*

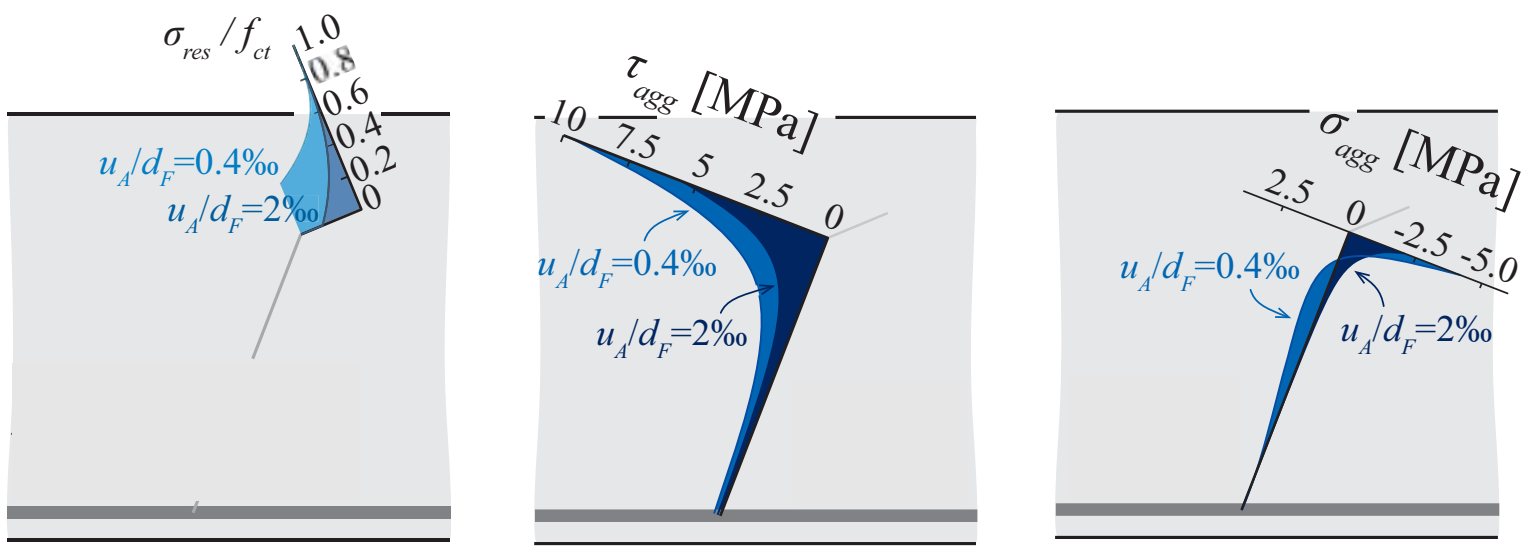

*Figure 2:* *Calculation of stresses for a given simplified kinematics to derive a physical model consistent with test observations (figure adapted from [2])*

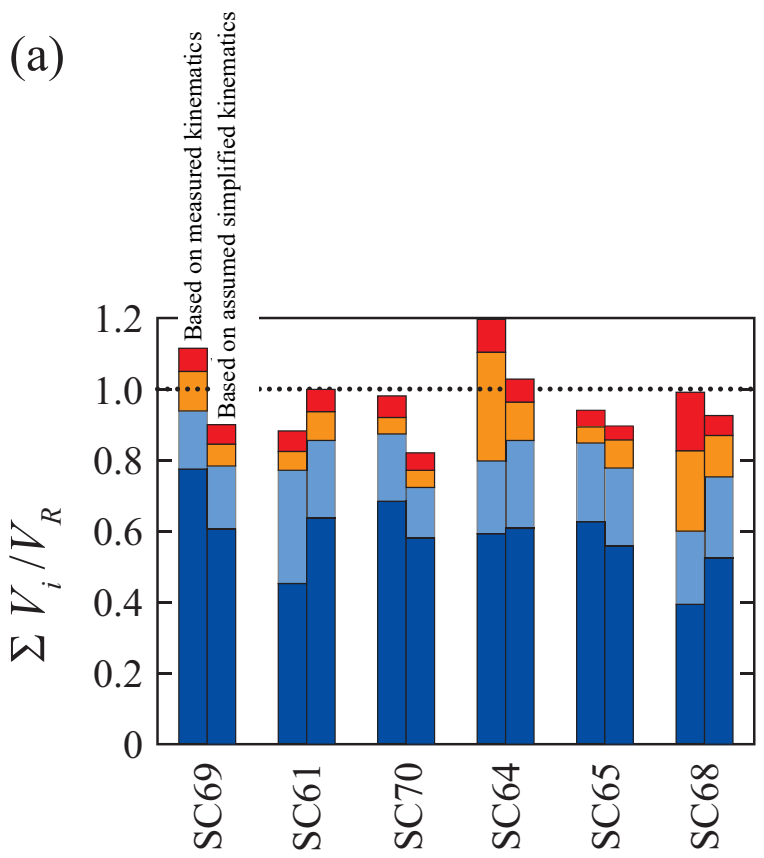

*Figure 3:* *Comparison of shear-transfer actions for simply supported beams (figure adapted from [8]): (a) contributions derived from tests (left hand side columns) and calculated according to the physical model (right hand side column)*

It can be noted with respect to this approach that interpretations of a phenomenon on the basis of test results is significantly more reliable and undisputable than based on FEM, as performed in the document by Dönmez and Bažant [1]. Blind competitions with specialized software calibrated for concrete shows the potential lack of consistency of this approach as well as large scatters depending on the software used or even in the manner a specific software is used.

**General considerations on the consistency of the CSCT**

The document produced by Dönmez and Bažant [1] questions some of the assumptions of the CSCT. The following general considerations are stated:

- Both the analytical formula proposed by Bažant to approximate the Size Effect Law (SEL, formula fitted by means of correlation to tests between two physical boundaries) and the formula for the SEL derived from the failure criterion of the CSCT (Eq. (7) for simply supported beams in [1], refer to original derivation in [7]) agree in the asymptotical response to limit cases for statically determinate beams in shear. Also, they agree in the shape (convex) of the transition between limit analysis for small sizes and Linear Elastic Fracture Mechanics (LEFM) for very large sizes.
- Size effect was a topic early investigated by the CSCT in a theoretical and experimental manner (see for instance [9]). It was however in 2015 when the full SEL was investigated [7]. The results proved the CSCT to be consistent to Bažant's approach, without the need of any change or additional consideration. This result confirms the consistency of the adopted failure criterion of the CSCT and of the simplified assumptions of the theory that suitably reproduces the phenomenon (although it was not originally calibrated to do so).
- The approach of the CSCT has also some fundamental advantages with respect to the approach used to characterize the SEL in shear problems by Bažant. These advantages are mainly justified by the fact that the theory allows considering both linear and nonlinear responses of reinforced concrete elements (an extensive discussion on this topic can be found in [10]):

- When the response of the element is linear (proportionality for a given element between the applied shear and crack openings) the slope of the asymptote governed by linear elastic fracture mechanics is consistently reproduced [7,10]. This is for instance the case of statically determinate beams (which refers to the classical academic test) and provides the slope -1/2 in a double-log plot (as stated in the document of Dönmez and Bažant [1]).
- Other than this rather academic case, suitably reproduced by both the CSCT and the SEL formula of Bažant, the CSCT approach has the advantage to reproduce nonlinear concrete responses (not constant ratio for a given element between the applied shear and crack openings). This is typically the case of continuous beams and slab where tension-stiffening may play a significant role (the shear/bending ratio varies during the loading process [10] ) or, as a very important case, of punching of slabs [10].

  With respect to punching, In the vicinity of slab-column connections, significant redistributions between radial and tangential moments occur due to concrete cracking and reinforcement yielding. This yields to a flexural response which is very different to a linear one. In these cases, by using a realistic response of the member (nonlinearly relating shear and flexural demands), the influence of size effect can be calculated in a tailored manner. In agreement to fracture mechanics, the size effect calculated by this approach yields to milder slopes of the size effect for asymptotically large sizes. Typical values are between -1/2.5 and -1/3 (punching, continuous beams [10]). Other researchers have reached the same conclusion with theoretical approaches (Broms [11], based on nonlinear fracture mechanics) as well as with experiments [12]. It is interesting to note that even Bažant clearly observed such a milder slope in punching when analysing test results [13], although he has not considered it for design and has accepted for these cases the (conservative) linear-elastic fracture mechanics slope of -1/2. In fact, the slope -1/2 is generally valid when linear-elastic behaviour is assumed as for uncracked responses, which is usually not relevant for practice.

- It is interesting to note that the CSCT is a consistent theory in its whole. The approach suggested by Bažant is actually based on a formula empirically calibrated to fit between two physical limit slopes of behaviour, while the CSCT formula describing the SEL is derived directly on the basis of its equations (failure criterion and flexural response). In addition, Bažant's proposal for SEL is plugged into an empirical expression to determine the shear strength without any further consideration. This raises questions on the consistency of the phenomena treated: is it applicable a multiplicative approach independently of all other parameters (potential nonlinearities, static systems…)
- It is the opinion of the authors of this document that it is possible to derive the size effect as a consequence of the mechanics governing shear failures and not imposing it (in a multiplicative manner) into a formula calibrated empirically on the results of academic beam tests. This latter approach (suggested by Bažant) yields to designers, scientists and engineers to a poor understanding of the phenomenon of shear failure and of its governing parameters (are larger crack widths associated to lower strengths?, does the slenderness play a role in it?, is the size effect behaviour the same for statically determinate beams than for redundant beams and slab-column connections?)

**Detailed reply to the criticism raised by Dönmez and Bažant**

In [1], criticism is raised on six aspects dealing with the CSCT. These criticisms are addressed below item by item:

- Re H1: The crack width is actually variable along the crack (see for instance Fig. 4a,b). This is the basis of the CSCT [2,7]. The response of the member is not calculated on the basis of a given crack width at a specific location, but on the basis of the crack opening and sliding profile along the whole critical shear crack. A detailed analysis on this topic and the consistent integration of the associated stresses can be consulted in fine details in [2,3,4,7,8]. The failure criterion accounts for various contributions [2,7]: residual tensile strength, aggregate interlock, dowelling action and inclination of compression chord.
- Re H2: It seems that Dönmez and Bažant have actually not understood that Eq. (1) refers to the integral response of the integration of stresses along the free-body defined by the critical shear crack and not to the aggregate interlock stresses (refer to previous reply). Dönmez and Bažant may refer to the aggregate interlock laws used in [2]
- Re H3-H4: As previously stated, the CSCT considers in its failure criterion that the crack width varies along the critical shear crack and so do the stresses [2,7]. Yet, as the centre of rotations is located approximately at the tip of the critical shear crack, the crack relative displacements can be characterized by one parameter (refer to the linear profile of the horizontal component of crack widths shown in Fig. 4b), as for instance the crack width at the level of the reinforcement [7] or others [9].
  It can be noted that the actual location, shape and complete kinematics of the CSC can be determined by means of the complete model of the CSCT [2]. However, determining a governing control section and reference fibre is a justified assumption (see [2,7]) which eases the use in view of developing design equations.
- Re H5: The linearity between the level of applied shear force and the crack openings (estimated from the flexural cracked deformations) is adopted as an assumption largely validated by test measurements [8,10] for statically determined beams (see Fig. 4c).
- Re H6: The aim of the CSCT is not to replace fracture mechanics. It is to provide a mechanical model consistent with test observations and that can be proven to be in agreement with test results and physical boundary responses. It reproduces the SEL and many other phenomena [2,9], as well as it allows for performing predictions of behaviour (slenderness effect…) beyond the available experimental data.

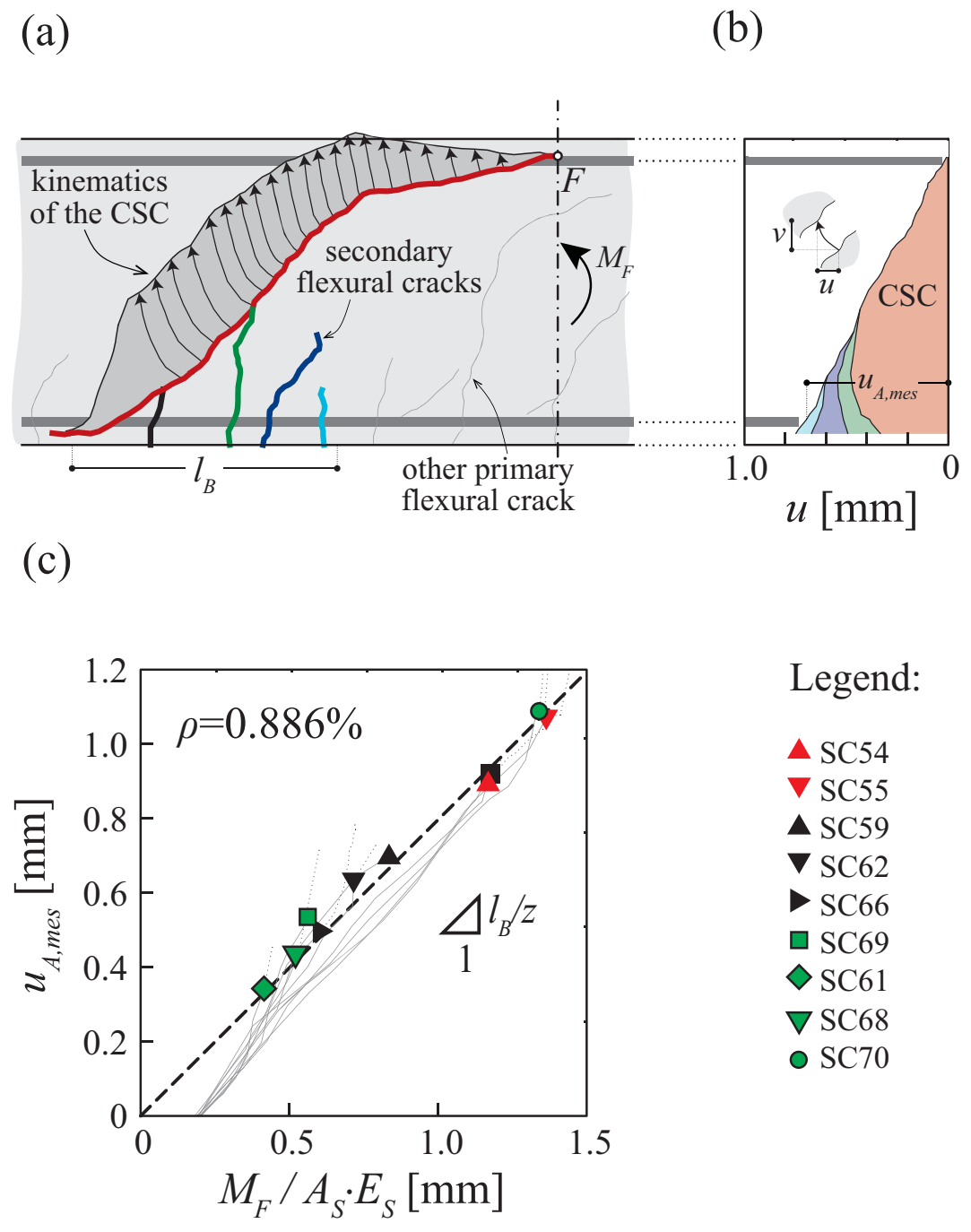

*Figure 4: Linearity between level of applied shear force and the crack openings (figure adapted from [8])*

**Concluding remarks:**

Once the reply to the criticism is given, the authors of this report would like to recall that the mechanics of the phenomenon of shear failures is complex. In order to yield to applicable design equations, simplifications on the mechanics shall be performed [2,7]. These simplifications may have a higher or lower level of accuracy depending on the case, but do not hide the mechanics beneath the phenomenon. On the contrary, they allow for a transparent interpretation of the phenomenon and understanding by students and designers. This is the approach that is currently being followed in *fib*'s Model Code as well as in the revision of Eurocode 2 (European design code for concrete construction).

In addition, the CSCT and similar approaches (as the Modified Compression Field Theory) allow developing design equations for shear and punching where it is assumed that the flexural reinforcement is dimensioned without any over-strength (LoA I in MC2010 for instance both for shear and punching). This is a very practical assumption for design, as the amount of flexural reinforcement is not necessary to be known when verifying the shear strength. For these cases, as demonstrated in [10], the influence of size on shear strength is stronger than for geometrically scaled specimens (with constant flexural reinforcement ratio) as not only the size, but also the amount of flexural reinforcement, are varied for increasing sizes (decreasing flexural reinforcement ratio with increasing sizes). In addition, the CSCT allows designing for cases where the flexural reinforcement is strained beyond the yield strain (due to redundancy). This is particularly relevant for estimating the deformation capacity in bending which is affected by plastic strains and size effects [14].

In a historical perspective, the work performed by prof. Bažant and his co-workers in the last decades is important as it has allowed clarifying the significance of size effect. Also, complying with the SEL can be considered as a benchmark for shear design models (slope -1/2 upon linear elastic responses, milder slopes in nonlinear cases). However, imposing the result of the SEL into an empirical equation for shear design and not deriving it from the mechanics beneath the phenomenon does not seem to be a consistent legacy for the future generations of structural engineers.